\documentclass[sigconf,authorversion,nonacm]{acmart}    

\usepackage{soul}

\AtBeginDocument{%
  }

\setcopyright{acmcopyright}
\copyrightyear{2023}
\acmYear{2023}
\acmDOI{XXXXXXX.XXXXXXX}

\acmConference[TEI '24]{Eighteenth International Conference on Tangible, Embedded, and Embodied Interaction}{February 11--14, 2024}{Cork, Ireland}
\acmPrice{15.00}
\acmISBN{978-1-4503-XXXX-X/18/06}

\begin{document}

\title{LensLeech: On-Lens Interaction for Arbitrary Camera Devices}

\author{Christopher Getschmann}
\email{cget@cs.aau.dk}
\orcid{0000-0002-0174-5974}
\affiliation{%
  \institution{Aalborg University}
  \city{Aalborg}
  \country{Denmark}
}

\author{Florian Echtler}
\email{floech@cs.aau.dk}
\affiliation{%
  \institution{Aalborg University}
  \city{Aalborg}
  \country{Denmark}
}

\renewcommand{\shortauthors}{}

\begin{abstract}

Cameras provide a vast amount of information at high rates and are part of many specialized or general-purpose devices. This versatility makes them suitable for many interaction scenarios, yet they are constrained by geometry and require objects to keep a minimum distance for focusing.
We present the LensLeech, a soft silicone cylinder that can be placed directly on or above lenses. The clear body itself acts as a lens to focus a marker pattern from its surface into the camera it sits on. This allows us to detect rotation, translation, and deformation-based gestures such as pressing or squeezing the soft silicone. 
We discuss design requirements, describe fabrication processes, and report on the limitations of such on-lens widgets.
To demonstrate the versatility of LensLeeches, we built prototypes to show application examples for wearable cameras, smartphones, and interchangeable-lens cameras, extending existing devices by providing both optical input and output for new functionality.

\end{abstract}



\begin{teaserfigure}
    \includegraphics[width=\textwidth]{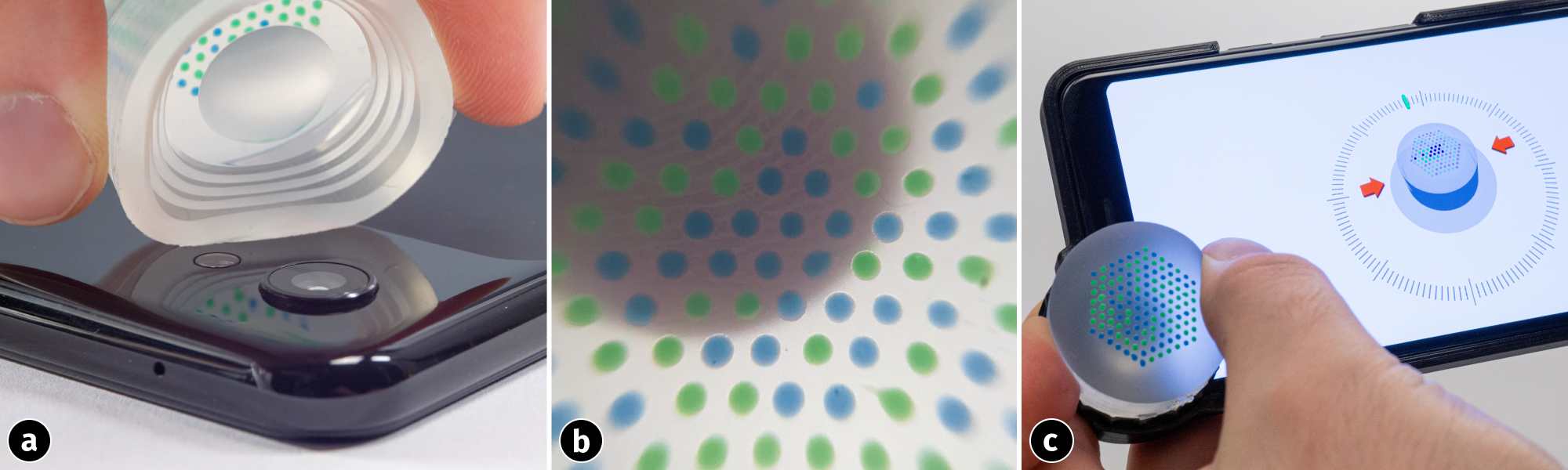}
    \caption{a) The soft silicone attachment with an integrated lens can be placed directly on and above cameras. b) A marker point pattern is focused by the silicone lens into the camera. c) Arbitrary devices with cameras, such as a smartphone can track the position, rotation, and deformation of the silicone attachment on the camera to sense input.}
    \label{fig:header}
\end{teaserfigure}

\maketitle

\section{Introduction}

There is an increasing number of mobile devices that make use of cameras as primary or additional sensors. At the same time, physical input has become a scarce feature on modern, highly-integrated devices.
Many of these camera devices are limited in their input channels and the interaction techniques they can offer due to trade-offs and design decisions to make them smaller, more robust, or less expensive.
Very small devices such as action cameras or wearables have either very few buttons or small touch screens, requiring to delegate even basic input tasks to paired smartphones or suffer from the fat finger problem ~\cite{siekFatFingerWorries2005}. 
Larger, interchangeable-lens cameras do provide both larger touchscreens and more buttons but could benefit from additional input options such as back-of-device interaction as well ~\cite{baudischBackofdeviceInteractionAllows2009}. 
Smartphones and tablets with capacitive touchscreens offer no physical feedback, have reachability issues ~\cite{wobbrockPerformanceHandPostures2008}, and require visual confirmation for input ~\cite{buxtonIssuesTechniquesTouchsensitive1985}.
While all of these devices could benefit from additional physical input, they have in common that they include a powerful sensor: a camera. Yet, any camera requires a minimal focal distance to provide focused images necessary to process rich user input ~\cite{xiaoLensGestureAugmentingMobile2013, yamadaCamTrackPointCameraBasedPointing2018}.

We present the LensLeech, a soft silicone attachment that can be placed directly on or above the front element of camera lenses. The silicone is optically clear and deformable so forces applied by fingers, hands, or arbitrary objects can be detected visually. By using the lower surface of the silicone body as a close-focus lens, a marker pattern on the opposite surface is always in focus regardless of the minimum focal distance of the camera lens.

With the LensLeech we can transform an unused or idle camera lens into a button, a knob, or a d-pad widget (and reverse it in seconds). These widgets provide physical feedback (when deformed), can be operated with gloves, and are robust, versatile, and inexpensive.
This allows to add knobs and d-pads to small wearable cameras to change settings in situ, make lens caps for large cameras touch-sensitive, or introduce novel optical attachments for smartphones.

In summary, we contribute:
\begin{itemize}
    \item a tangible deformation sensor to create buttons, knobs, and d-pads, combining soft body, optical elements, and sensing pattern in a single object
    \item discussions on the design and fabrication of on-lens widgets
    \item an image processing pipeline for analyzing position, rotation, and deformation
    \item application examples for integration with new and existing devices
\end{itemize}

Many research approaches aim at providing novel functionality with new or existing sensors for future devices, often built on the assumption or requirement of a possible miniaturization and integration of this external sensing hardware into a new device with a new form factor. However, we explicitly aim at retrofitting existing and well-proven interaction techniques to sensors that make them available both to legacy devices today as well as new ones in the future. This could help to extend the lifetime of devices in circulation by improving their usability and reducing incentives to update to newer hardware prematurely. 
A user study (with regard to the input capabilities of the widgets) is not presented as these input modalities are well understood and can be directly applied to this new form factor.

The remainder of this paper is organized as follows: related work is discussed with a broad overview of vision-based elastomer sensors and on-lens/around-lens interaction techniques, then we explain our concept of soft silicone attachments for on-lens interaction sensing. The image processing pipeline and fabrication procedure are summarized subsequently. We built upon that by presenting a set of scenarios and prototypes created to show real-world applications. Finally, we discuss the limitations of using soft silicone attachments for on-lens interaction and conclude with specific directions for future work.

\section{Related Work}

Relevant to the presented work are both optical deformation sensors primarily developed for robotic applications as well as human-computer interaction techniques and prototypes that gather input from the space on and around camera lenses.

\subsection{Optical Elastomer Sensors}

Optical deformation sensing of soft materials is performed either by measuring light altered by the surface or by detecting displacement of high-contrast markers, on the surface or encapsulated in the material.
Surface deformation measurements have been proposed based on total internal reflection ~\cite{hiraishiObjectProfileDetection1988}, Lambertian reflection ~\cite{johnsonRetrographicSensingMeasurement2009, watanabeGenericMethodCrafting2014, dongImprovedGelSightTactile2017, donlonGelSlimHighResolutionCompact2018, taylorGelSlim3HighResolutionMeasurement2021, wangGelSightWedgeMeasuring2021} and polarization ~\cite{satoPhotoelasticTouchTransparentRubbery2009}.
The most common type of sensor, the \textit{GelSight} family, makes use of Lambertian reflection by coating the clear elastomer with a reflective membrane. Multispectral illumination from below allows to derive deformation depth and thus a detailed 2.5d geometry of the reflective surface.
For marker-based sensing, high-contrast points are painted on the clear surface of the elastomer ~\cite{wangGelSightWedgeMeasuring2021, taylorGelSlim3HighResolutionMeasurement2021, dongImprovedGelSightTactile2017}, on the interior of an opaque hull for TacTip sensors ~\cite{winstoneTACTIPTactileFingertip2012, ward-cherrierTacTipFamilySoft2018} or colored balls are directly encapsulated in the soft material ~\cite{kamiyamaEvaluationVisionbasedTactile2004, sferrazzaDesignMotivationEvaluation2019, yamaguchiImplementingTactileBehaviors2017}.

These sensors have been used extensively for tactile sensing in robotic applications, mounting the sensor on the end effector to measure gripping force and detect slipping. For this, the sensor assembly is designed as a monolithic unit consisting of camera sensor, lens, and elastomer block. While mirrors ~\cite{donlonGelSlimHighResolutionCompact2018, wangGelSightWedgeMeasuring2021} and fisheye lenses ~\cite{taylorGelSlim3HighResolutionMeasurement2021} have been used to shorten optical paths to create more compact grippers these sensors are still of considerable size and rely on a tight integration of all components, making them incompatible with arbitrary cameras. Modular approaches offer only exchangeable elastomers while still using a specialized camera ~\cite{lambetaDIGITNovelDesign2020}. Additionally, all gel-based sensors with the exception of the sensor by Obinata et al. ~\cite{obinataVisionBasedTactile2007} and \textit{Fingervision} ~\cite{yamaguchiImplementingTactileBehaviors2017} block environmental light and require white, RGB or ultraviolet illumination by integrated LEDs. 
For a detailed overview refer to the reviews by Shimonomura ~\cite{shimonomuraTactileImageSensors2019} and Abad et al. ~\cite{abadVisuotactileSensorsEmphasis2020}.

In the domain of human-computer interaction, elastomer sensors have been used for interactive surfaces ~\cite{follmerDeFormInteractiveMalleable2011}, clay-like projection displays ~\cite{punpongsanonDeforMeProjectionbasedVisualization2013} and tangibles on tabletops ~\cite{weissSLAPWidgetsBridging2009, henneckeOpticalPressureSensing2011a} to support novel interaction techniques.

\subsection{On-Lens/Around-Lens Interaction}

Placing a fingertip directly on a smartphone camera lens has been proposed as an interaction technique in \textit{LensGestures} ~\cite{xiaoLensGestureAugmentingMobile2013}. The unfocused environmental light passing through a finger's tissue is used to approximate finger positions and recognize gestures.
\textit{CamTrackPoint} ~\cite{yamadaCamTrackPointCameraBasedPointing2018} improves on this concept by providing tactile feedback. A spring-actuated plastic ring is integrated with a smartphone case directly over the lens for the finger to rest on. The thin ring blocks light with a sharp transition to black and provides a higher precision compared to tracking the blurred finger.
A proof-of-concept for more complex on-lens input techniques is presented by Watanabe et al. ~\cite{watanabeGenericMethodCrafting2014}: soft and optically clear toys with a reflective surface coating are placed on the camera while a neural network is trained to recognize deformation/gestures from internal reflections observed through a hole in the bottom.
This represents the simplest and most basic on-lens widget: unfocused, untagged, unpowered and depending on natural illumination, but very easy to manufacture and not obstructing the camera when not in use.

Interaction in the space around lenses requires mirrors to both shorten the optical path and redirect light. \textit{Clipwidgets} ~\cite{visschedijkClipWidgets3DprintedModular2022} makes use of a conical mirror in a bulky smartphone case to read the state of physical widgets such as buttons and sliders. Similar approaches have been presented for back-of-device interaction concepts with smartphones ~\cite{wongBackMirrorBackofdeviceOnehanded2016, matsushimaAttachingObjectsSmartphones2017, kitadePrismoduleModularUI2019}.
Without relying on physical input objects \textit{Handsee} ~\cite{yuHandSeeEnablingFull2019} utilizes a prism to track hands touching and floating above a smartphone display while \textit{Surroundsee} ~\cite{yangSurroundseeEnablingPeripheral2013} tracks objects in the whole room with a circular 360-degree mirror above the smartphone camera.

Similar techniques have been used without mirrors or lenses in the context of tangibles with silicone feet for pressure sensing ~\cite{weissSLAPWidgetsBridging2009}, deformation sensing on small wearables ~\cite{weigelDeformWearDeformationInput2017}, and surface position sensing with fibers ~\cite{wimmerFlyEyeGraspsensitiveSurfaces2010}.
Other work that is related to the presented concept is \textit{Bokode}~\cite{mohanBokodeImperceptibleVisual2009}, a marker made of a lenslet and microfilm which magnifies a grid of 2D barcodes into the defocused lens of a camera and \textit{Sauron}~\cite{savageSauronEmbeddedSinglecamera2013a}, a design tool to integrate cameras in hollow objects that read the state of mechanical input elements. 

While physical input similar to the LensLeech can be achieved on smartphones in particular by simply redirecting electrodes of the capacitive touchscreen ~\cite{yuCliponGadgetsExpanding2011, schmitzFlexiblesDeformationAware3DPrinted2017a, matsushimaAttachingObjectsSmartphones2017}, the LensLeech is not limited to touchscreens and can be applied across a range of devices, see the application examples in section \ref{section:applicationexamples}.

Since on-lens interaction concepts such as \textit{CamTrackPoint} and \textit{LensGestures} make use of unfocused light, they are limited in their expressiveness due to the low amount of information available. While they are suitable for smartphones and their scratch-resistant camera assemblies, these concepts translate poorly to interchangeable lens cameras or action cams with lens front elements often using coatings sensitive to scratches or prints from fingertips.
This is one of the fundamental issues we intend to address with our generalizable approach.

\section{The design of on-lens widgets}

\begin{figure}
    \centering
    \includegraphics[width=\linewidth]{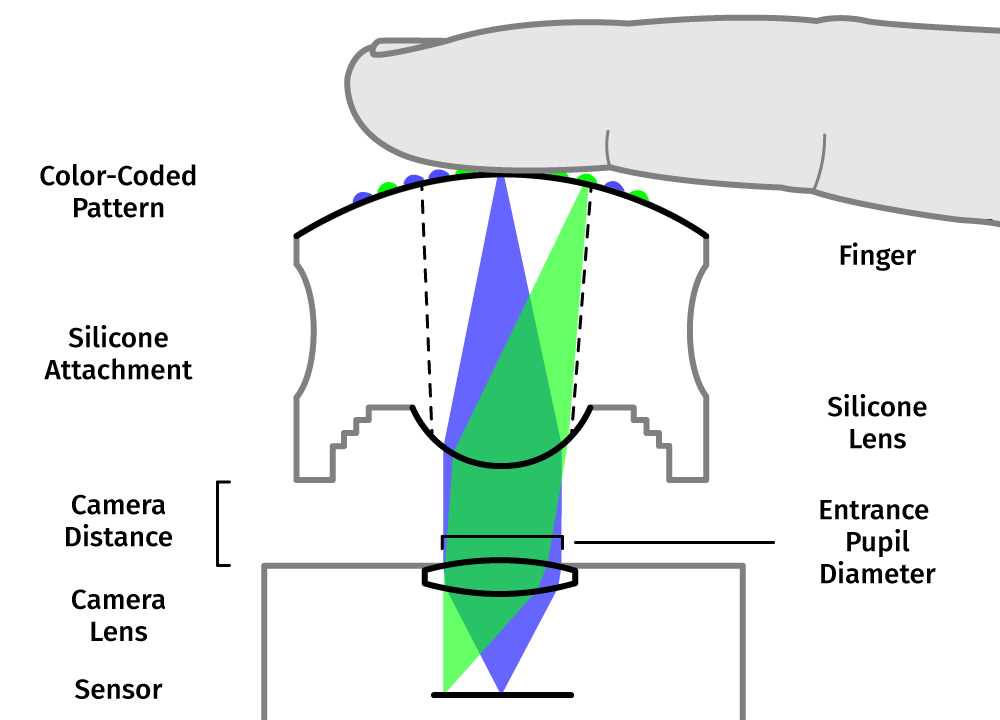}
    \caption{Illustrative ray diagram of the combined optical system. The field of view inside the silicone (dashed line) depends on the field of view of the camera, the position and diameter of the entrance pupil, as well as the distance between silicone and camera lens.}
    \label{fig:raydiagram}
\end{figure}

We propose that any physical attachment enabling on or around-lens interaction with both existing and future devices should---ideally---adhere to these basic design considerations:

\begin{itemize}
    \item \textbf{safe} to use near or on optical components and providing credible reassurance to the user about this. This is a prerequisite for user acceptance.
    \item \textbf{non-invasive}, requiring no hardware modifications of the host device or its camera. This ensures compatibility with existing devices that benefit most from optical attachments.
    \item \textbf{passive} and unpowered, requiring only ambient illumination (if possible) to reduce size and complexity.
    \item \textbf{universal}; compatible with arbitrary camera/lens combinations across a wide range of device types.
\end{itemize}

Elastomer sensors in general fulfill the first and most important of these requirements by virtue of their nature: they are soft. However, existing sensors fall short in most or all other points.

As discussed, these sensors combine camera and elastomer in a permanent assembly with a fixed position and rotation, limiting the way objects can interact with them. Additionally, they make use of known sensor and lens combinations to allow camera calibration and optimizations of sensor geometry (for example by backprojecting through a calibrated lens to find optimal marker point placements). This makes these sensors more precise and reliable but prevents them from being used with arbitrary lenses and cameras.
Finally, most sensors require constant internal illumination. Reflective membranes (GelSight) and rubber skins (TacTip) are blocking ambient light to avoid interference. Only sensors relying solely on point patterns ~\cite{obinataVisionBasedTactile2007, yamaguchiImplementingTactileBehaviors2017} can tolerate ambient illumination.

We propose an elastomer sensor design suitable for interaction sensing. The LensLeech is a tangible soft input device that resembles the gel part of an elastomer sensor. Our all-silicone design combines a lens, compliant body, and a colored marker pattern in a single unit (see fig. ~\ref{fig:raydiagram}). This addresses all design requirements at the cost of reduced reliability and precision compared to elastomer sensor assemblies that are designed to measure precise gripping forces on robot actuators.

The small form factor of the LensLeech attachment (33mm diameter, 25.5mm height) makes it easy to grip it with two fingers and place it on a camera. By using the lower surface of the clear silicone body as a lens, light reflected by the deformation-sensing pattern on the surface is collimated and can be focused on the sensor at any distance from the camera. This makes it possible to place the silicone foot of the LensLeech directly on or slightly above the front element of a wide range of lenses.
The combined optical system of sensor, camera lens, silicone lens, and deformation sensing pattern is limited by the field of view of the camera, its entrance pupil, and the distance to the silicone attachment. This is discussed in more detail in section \ref{section:limitations} (Limitations).

\hfill\break
\noindent\textbf{Marker Pattern}\newline
When choosing a marker pattern for deformation sensing, we need to take into account that a positive lens required to move the focal point to the surface of the silicone body will introduce a strong magnification effect. This amplifies any defects or irregularities in the pattern and requires the fabrication of very small features. The most precise and reliable method is the deposition of single droplets of silicone paint. This makes a point pattern the preferred choice.

A point pattern is used by other optical tactile sensors such as TacTip and GelSight as well, however, these sensors are fixed assemblies that can compute deviations from a static reference frame. This does not apply to a silicone sensor that can be moved and rotated freely, thus a method to align the currently visible region of interest with the overall marker grid is required. 

A common method for identifying sections of point grids are two-dimensional DeBruijn sequences. These are sequences that contain every subsequence of a defined size at most once. Printed as microdots on paper these have been used for position-tracking with digital pens ~\cite{anoto} (encoding bits as a displacement from a regular grid) and tangibles ~\cite{schusselbauerDothrakiTrackingTangibles2021} (encoding bits as black and white). However, unlike a rigid piece of paper which allows displacement coding of dots, the soft silicone is easily bent or compressed and requires coding by color or contrast. 

\begin{figure}
    \centering
    \includegraphics[width=\linewidth]{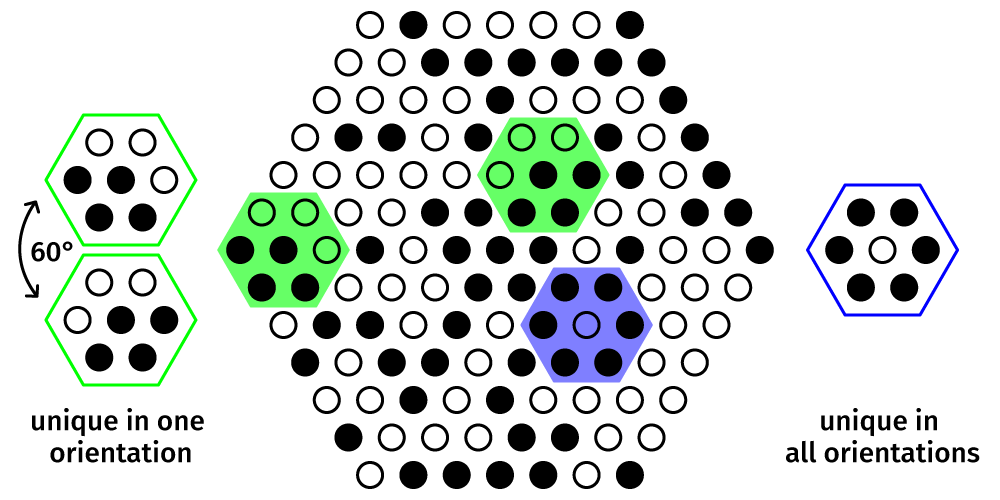}
    \caption{Each hexagonal sliding window appears only once. Some sliding windows are unique in all six orientations, some can be found in multiple locations when rotated.}
    \label{fig:pattern}
\end{figure}

While a hexagonal arrangement of points offers the densest packing it is incompatible with a 2D-DeBruijn sequence. Hence, we computed a DeBruijn-like pattern with 7-point hexagons instead of 3x3 matrices using a brute-force approach. Each overlapping hexagonal sliding window in the pattern is unique in the given rotation (see fig. \ref{fig:pattern}). An optimal pattern does contain only hexagons that are unique in all rotations which is simplifying pattern matching during image processing, however, this requires a minimum of three colors at a suitable pattern size. Higher robustness to adverse lighting conditions and a less error-prone fabrication with only two different paints is the reason why a less-than-optimal two-color pattern is preferable.

Our hexagon patterns consist of 127 points and require 91 unique sliding windows. This is sufficient to cover the visible region of the silicone attachment even when placed on a wide-angle camera. While other sensors such as TacTip and GelSight Wedge use the same or similar number of points, only a subset of points is visible at a time for our application due to the magnification of the silicone lens. If a unique center hexagon is enforced during pattern generation up to 28 patterns can be discerned. This allows to map different silicone attachments to specific input modalities. %

\hfill\break
\noindent\textbf{Image Processing}\newline
The DeBruijn-like point pattern is color-coded in blue and green to offer a high contrast across the range of human skin tones. Coincidentally, the fingertips have fewer variations due to smaller differences in the skin tones of palms overall. The detection and classification of the points is the first step of the image processing pipeline. Background removal is performed by thresholding in the HSV color space. The diffuse top surface of the silicone body improves this step considerably without blocking any ambient light.
From these point candidates, colors are extracted and classified by thresholding the two classes in the hue component of the HSV colorspace using Otsu's method~\cite{otsuThresholdSelectionMethod1979}. This is robust to errors in white balance caused by tinted ambient illumination or fingertips and computationally less expensive than other classification methods. Robustness is especially important since most auto white balance algorithms overshoot for several dozen frames when a finger is placed on the point pattern. 

For pattern matching each detected point is grouped with its 6 closest neighbors and all 6 rotation variants are checked against a lookup table. Correct rotation is assumed when the highest number of matches is found between neighboring sliding windows in the camera image and ground truth pattern. Given an optimal pattern, only one rotation would result in a match (in the absence of any errors), yet this computationally-expensive step is necessary to limit the pattern to only two colors. This makes the pipeline's processing speed highly dependent on the number of detected points. On a laptop computer (2.3 GHz 8-Core Intel i9) 34 frames per second are processed when 39 points are visible (30\% of the pattern) and 21 FPS when 69 points (70\%) are visible. Smartphone performance numbers are not reported since the Android implementation performs segmentation on-device but outsources the pattern-matching step of the pipeline to a server.

%

\begin{figure}
    \centering
    \includegraphics[width=\linewidth]{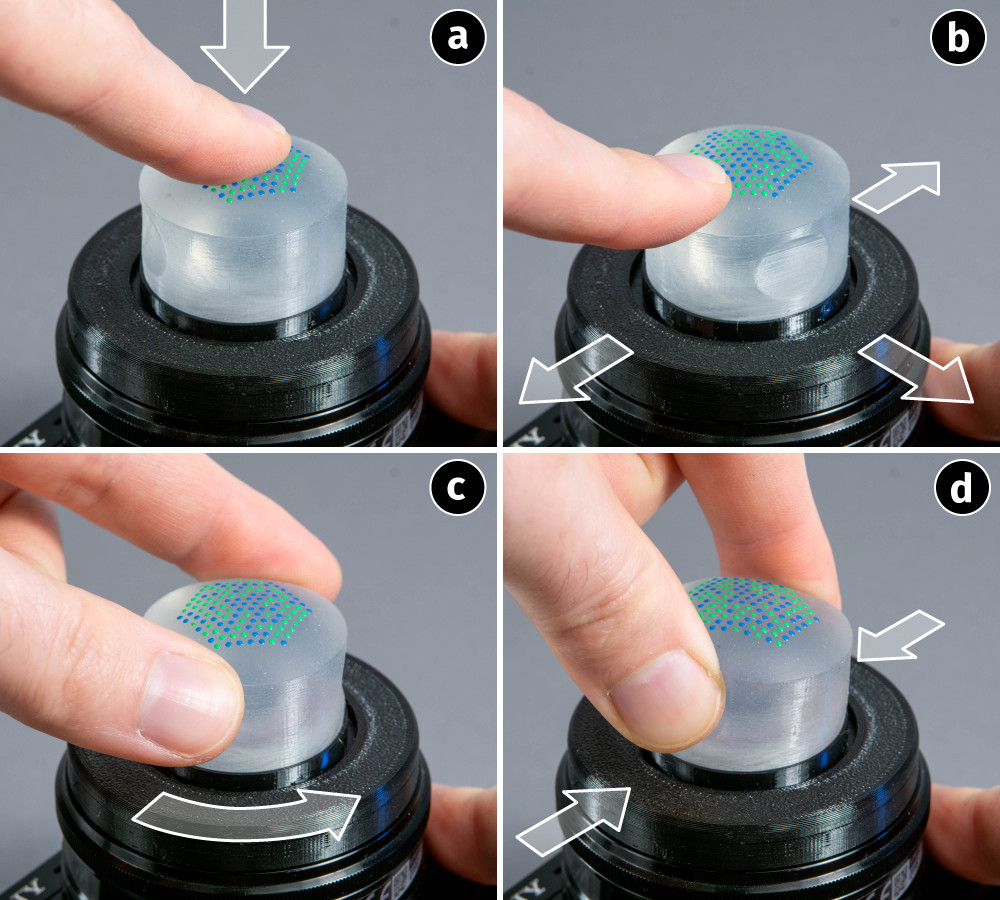}
    \caption{The four types of input: a) Pressing on the silicone body b) lateral pushing in any direction c) rotation on the optical axis d) squeezing the silicone.}
    \label{fig:interaction}
\end{figure}

The input gestures are derived directly from the matched point pattern (see fig. \ref{fig:interaction}). A press on the top is recognized by detecting locally increased distances between neighboring points, pushing sideways by computing the centroid of all detected points, rotation by Kabsch`s algorithm ~\cite{kabschSolutionBestRotation1976}, and squeeze by a global change of point distances along the squeeze axis. The gesture detection relies on algorithm implementations in SciPy~\cite{2020SciPy-NMeth}, while processing of image data is done using OpenCV~\cite{opencv_library}.

Before discussing how these input types inform examples for real-world application, the fabrication process of both silicone body and color-coding pattern is described briefly.

\section{Fabrication}

\begin{figure}
    \centering
    \includegraphics[width=\linewidth]{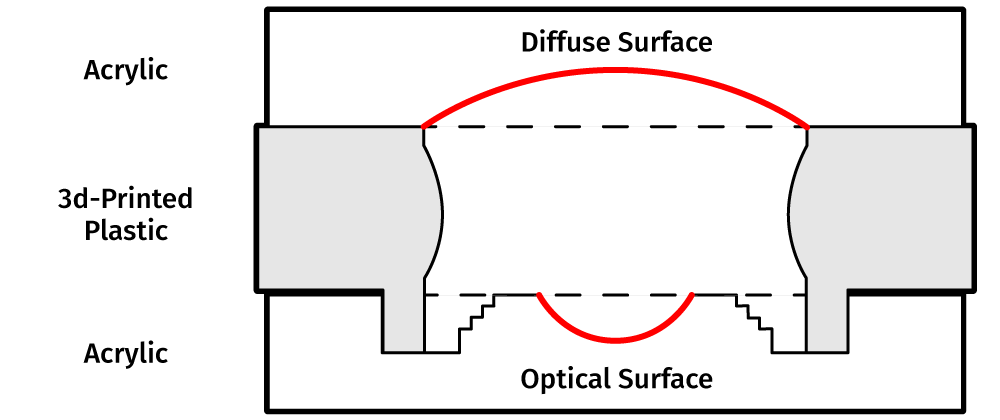}
    \caption{Cross section of the mold. The curved surfaces are ground and polished each with a precision steel ball of the required curvature. The optical surface is polished to a 2-micron finish and the diffuse surface to 40 microns. All three sections of the mold are aligned with metal dowel pins (not pictured). Liquid silicone is poured through a horizontal channel in the 3d-printed plastic part.}
    \label{fig:mold}
\end{figure}

\begin{figure}
    \centering
    \includegraphics[width=\linewidth]{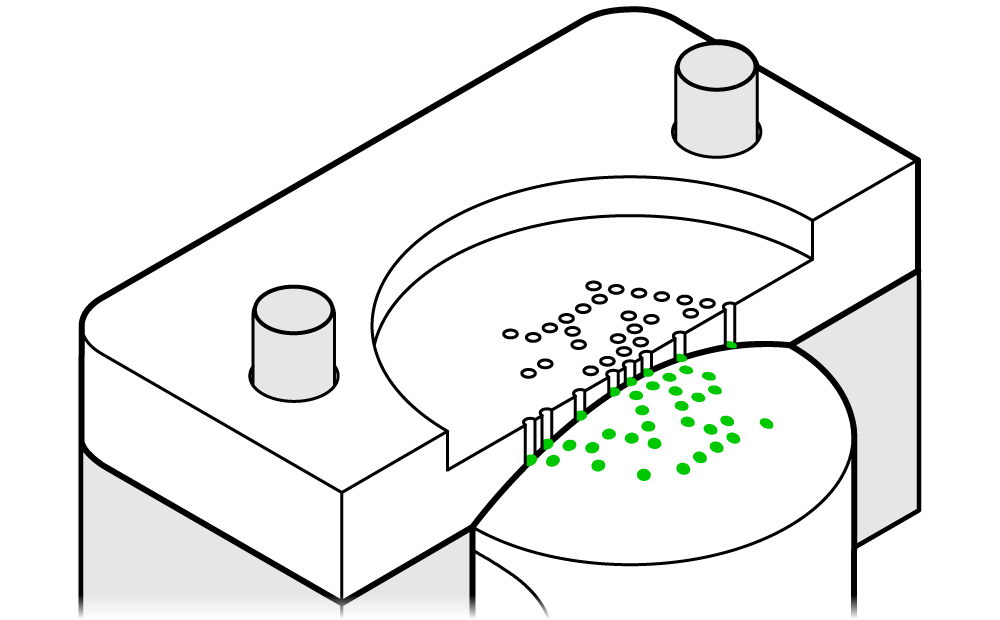}
    \caption{Cross section of the stenciling fixture. The silicone body is pressed upwards against the curved surface to create a seal. Once locked in place by clamps, the liquid pigmented silicone is poured into the recess at the top of the stencil and makes its way through the micro-drilled channels. When the stencil is lifted a small domed blob of partially-cured silicone paint remains on the surface. The stencil and fixture for the silicone body are aligned with metal dowel pins to ensure precise placement for each consecutive stencil and color.}
    \label{fig:stencil}
\end{figure}

The clear silicone body is created by mixing, degassing, and pouring liquid silicone (Trollfactory Type 19) into a mold and letting it cure.
The mold itself requires two precisely manufactured features. The lower cavity is an optical surface (sufficiently smooth to refract light for imaging applications) to create the spherical convex lens of 7.5mm radius for focusing.
The curved top surface of 30mm radius diffuses light. Both surfaces are CNC-milled from acrylic before being ground and polished. For this, the spherical surface of the acrylic part is coated with lapping paste and pressed against a rotating steel ball of matching radius (widely available as high-precision replacement parts for large ball bearings). %
After polishing the acrylic plates are fastened to a 3d-printed center part to complete the mold (see fig. \ref{fig:mold}).
Once cured and de-molded the point pattern is applied to the clear silicone body with two 3d-stencils milled from acrylic (one per color). The stencil is fabricated by drilling a duplicate of the mold top part with a circuit board drill (1.0mm) to create channels. The soft body is pressed into the matching cavity of the stencil from below and the pigmented silicone can be poured on the channels (see fig. \ref{fig:stencil}) before removing the remaining air from the channels in a vacuum chamber. The stencil guarantees correct placement and uniform point size. Only silicone itself bonds reliably to cured silicone parts, thus uncured silicone mixed with color pigments is the most suitable paint. The main issues in this process are ensuring that the high-viscosity silicone reliably fills the channels and avoiding oversaturation of the silicone with pigments in a silicone oil solution, which may inhibit the curing process. A mixture (by weight) of Smooth-On's Psycho Paint silicone with 15 percent dry UV-reactive pigment powder and 25 percent solvent (toluene) to lower the viscosity worked best. After 24 hours the pigmented silicone binds reliably to the optically-clear silicone body, creating a point pattern on the surface that is flexible and wear-resistant.

The choice of lens curvature during mold production is a trade-off. A mold for a lens with a stronger curvature is more demanding in fabrication but the shorter focal length allows to reduce the height of the silicone body. At the same time, it decreases the depth of field and the field of view, allowing to track a lower number of points in the pattern. 
Additionally, interacting with the LensLeech deforms both top surface and lens. A strong press on the top will reduce the height of the body by several millimeters depending on the hardness of the silicone. 
If the height of the silicone body does not match the focal length the light will not exit the system collimated, resulting in a pattern that would be out of focus. In reality, this is rarely an issue since autofocus cameras can compensate for this, fixed-focus cameras often have a sufficient depth of field, and the image processing pipeline is robust to low levels of blurring. 

A paraxial approximation of the focal length can be obtained using the lensmaker's equation. 
Only the refraction of the first surface is relevant for the LensLeech geometry, so a thin, plano-convex lens in air (\( d = 0, R_2 = \infty \)) can be assumed:

\[ \frac{1}{f} = (n-1) \left(\frac{1}{R_1}-\frac{1}{R_2}+\frac{(n-1)d}{nR_1R_2} \right) = \frac{n}{R_1} - \frac{1}{R_1} \]
\newline
We have chosen a radius of \( R_1 = 7.5mm \) for the lens surface and assume that Trollfactory Type 19 has a refractive index of \( n = 1.41 \), similar to other platinum-cure silicones.
This would result in a total focal length of 18.29mm. To account for the deformation during interaction and the strong curvature of the lens we increase the height of the silicone body by a factor of \( 1.3 \) to 25mm. While the LensLeech should be able to touch the glass surface of the camera lens, the silicone lens surface in the center requires an air gap to refract light. Thus we extend the foot around the lens by 1.0mm to account for deformations (cross section can be seen in fig. \ref{fig:raydiagram}). This provided the most reliable results during testing for high forces when pressing and squeezing while still keeping the point pattern well within the depth of field of most cameras when not deformed.

While fabrication is the primary challenge to making close-focus silicone lens attachments, a thorough description would go beyond the scope of this paper. Please refer to the companion repository\footnote{\url{https://github.com/volzotan/LensLeech}} for detailed information about the fabrication process.

\section{Application Examples}\label{section:applicationexamples}

Tactile on-lens input can be used in a variety of scenarios for devices with cameras in many sizes.
We present two application examples that show how on-lens interaction can be utilized to make input on both large and small cameras more convenient and transfer well-established tangible interaction techniques to smartphones.

\begin{figure}
    \centering
    \includegraphics[width=\linewidth]{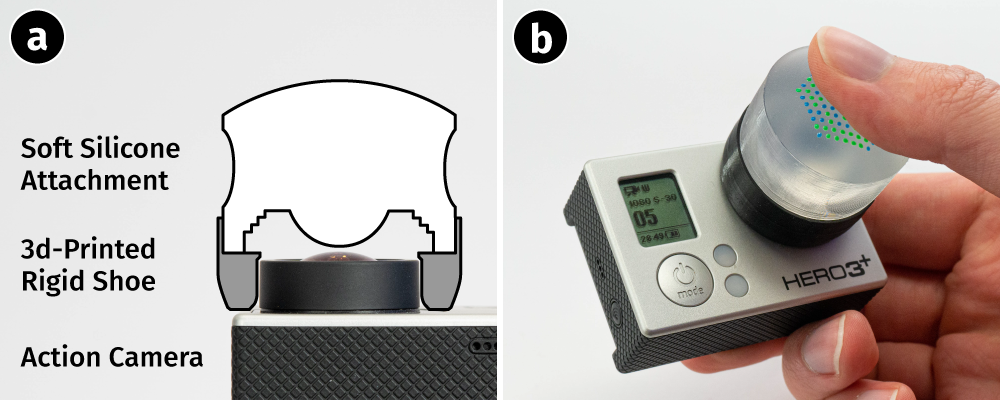}
    \caption{a) Cross section: the silicone attachment can be extended with a 3d-printed shoe matched to the specific device so it slides over the protruding wide-angle lens of an action camera. b) The LensLeech could be used as a rotating knob, press to confirm, squeeze to cancel.}
    \label{fig:actionknob}
\end{figure}

\hfill\break
\noindent\textbf{Interactive Lens Caps for Digital Cameras}\newline
Small action cameras offer a very limited number of buttons, a tiny display (if any), and only optionally, a touch interface on this display. While there are techniques to facilitate touch input on very small displays ~\cite{baudischBackofdeviceInteractionAllows2009}, it may be cumbersome. Adjusting settings is often performed via a companion app on a smartphone which is paired with the wearable camera. 
However, there are scenarios in which the phone is unavailable and direct interaction with the device itself may be favorable. These can be casual, everyday situations like wearing gloves or very specific use cases such as interacting with a camera enclosed in a waterproof housing while swimming or diving.
By adopting a lens cap or protective storage case that integrates a LensLeech (see fig. ~\ref{fig:actionknob}), we can add tangible controls such as a rotation knob or a d-pad, and extend the number of buttons on the device for easier navigation through nested menus. However, note that in the case of underwater usage when the LensLeech is pressed against a waterproof housing, a different silicone lens curvature will be required due to the refractive index of water.

\begin{figure}
    \centering
    \includegraphics[width=\linewidth]{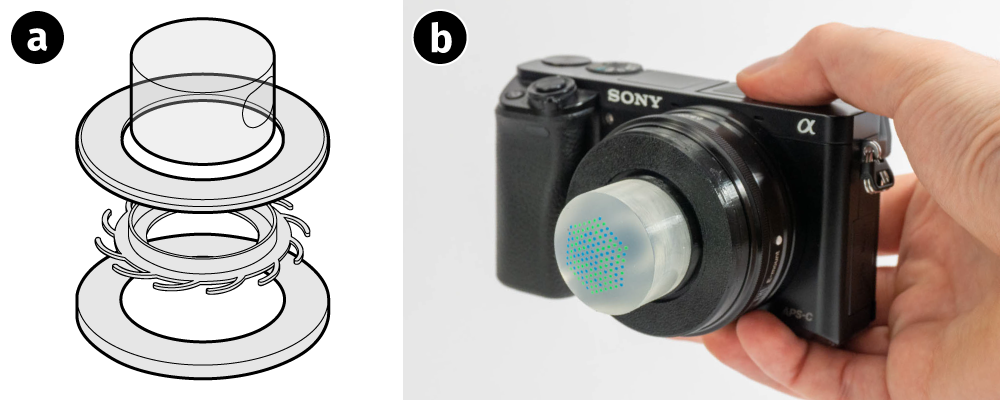}
    \caption{a) The silicone attachment is placed in a 3d-printed lens cap with a spring-loaded mechanism to allow lateral movement and rotation. b) The lens cap on an interchangeable-lens camera.}
    \label{fig:touchcap}
\end{figure}

This concept extends to larger cameras as well. Many digital consumer cameras are infamous for convoluted menus and poor usability in general. Browsing recorded videos and photos, changing settings in nested menus, and entering credentials to set up wireless connections requires prolonged attention and interaction while the sensor itself is not in use during these tasks. 
By integrating a silicone attachment into a lens cap we can leverage the unused hardware without interfering with the primary use case of the camera. Rotating or pushing the silicone sideways could be used to traverse large lists of settings or captured footage (see fig. ~\ref{fig:touchcap}). 
While the soft silicone can touch lens coatings without damaging them, the lens cap in this case prevents direct contact and may provide reassurance to the user for these very expensive lenses.

\begin{figure*}
    \centering
    \includegraphics[width=\textwidth]{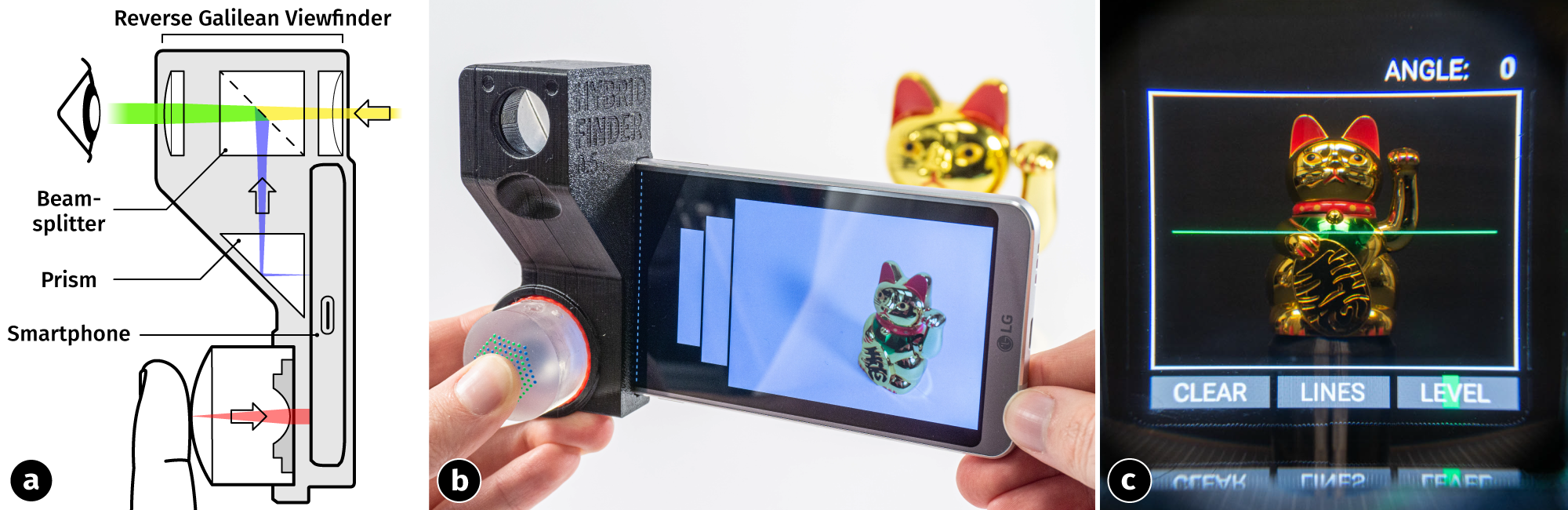}
    \caption{a) Cross section of the hybrid viewfinder. The beamsplitter overlays the light emitted by a section of the smartphone screen (blue) over the viewfinder image of the world (yellow), while the silicone attachment rests on the front camera (red). b) The finder can be slid over the top section of the smartphone. The silicone attachment provides rich tangible input to control settings and take a photo without visual confirmation as a touchscreen would require. c) View through the finder showing an overlay of the selection menu and a digital spirit level.}
    \label{fig:hybridviewfinder}
\end{figure*}

\hfill\break
\noindent\textbf{Hybrid Viewfinders for Smartphones}\newline
While the LensLeech provides optical input to camera-based devices, it can be combined with other components that offer optical output as well. This allows to create complex passive optical add-ons to existing devices. 
The hybrid viewfinder slides over the top section of a smartphone covering the front lens and a portion of the display. By adding a beamsplitter prism to a camera viewfinder, a section of the covered display can be reflected into the viewfinder's optical path (see fig. ~\ref{fig:hybridviewfinder}a) to create a hybrid optical/electronic viewfinder for smartphone photography. By integrating the LensLeech into the viewfinder attachment, the front camera can be used for input while the rear camera takes images and optionally provides data for the viewfinder overlay (see fig. ~\ref{fig:hybridviewfinder}c). 
Rotating the LensLeech changes the data overlay and pressing it triggers image capture. This allows optical input and output with no hardware modifications, transforming a smartphone into a modern rangefinder-style camera.

\section{Evaluation \& Limitations}\label{section:limitations}

\begin{figure}
    \centering
    \includegraphics[width=\linewidth]{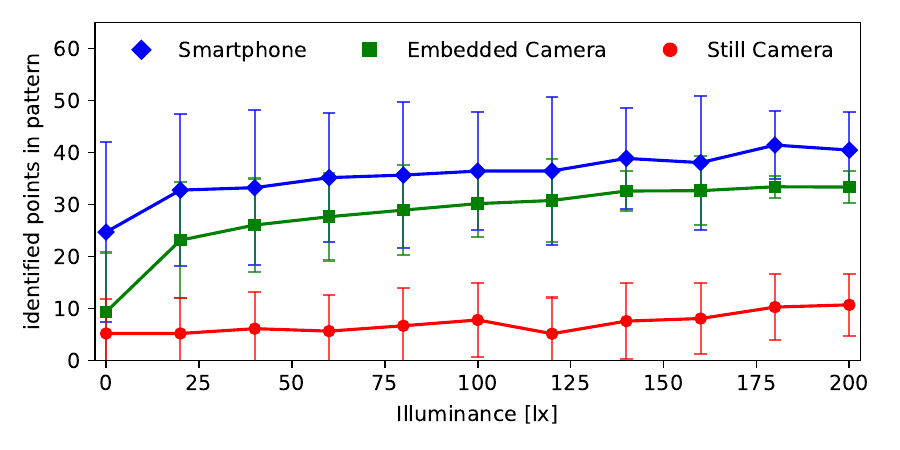}
    \caption{Mean number of identified points in relation to ambient illumination strength. Error bars specify standard deviation. Monochromatic or environmental light with a strong color tint will result in an undetectable point pattern regardless of illumination strength, this is the main cause for outliers in the plot. Note: the maximum number of points visible to the camera varies across devices depending on pupil size and field of view.}
    \label{fig:plot_patternvslight}
\end{figure}

When tested with artificially generated images (rendered images of the deformation point pattern with a pinhole aperture instead of a lens for focusing) the rotational error is negligible at an average of 0.03 degrees. More relevant and considerably more challenging is the real-world performance under low-light conditions and ambient light with color casts. As an evaluation setup, three cameras with an attached LensLeech (Pixel 3a smartphone, Sony A6000 + Sony 20mm 2.8 digital still camera, Raspberry Pi V1 embedded camera) were placed in complete darkness facing a display showing a subset (201 images, three per category) of the MIT indoor scene recognition dataset ~\cite{quattoniRecognizingIndoorScenes2009}. These were artificially darkened and brightened to simulate a low-light environment (resulting in a total of 1206 images). The illuminance of each scene was measured at the surface of the LensLeech with a TSL2591 ambient light sensor.
Detection performance is depending on the combination of sensor, lens, and environment, but in general, it can be observed that above 150 lux reliable operation can be expected (see fig.\ref{fig:plot_patternvslight}). Indoor lighting conditions usually exceed 150 lux while 300-500 lux are recommended for office work ~\cite{ISO8995}. %

\begin{figure}
    \centering
    \includegraphics[width=\linewidth]{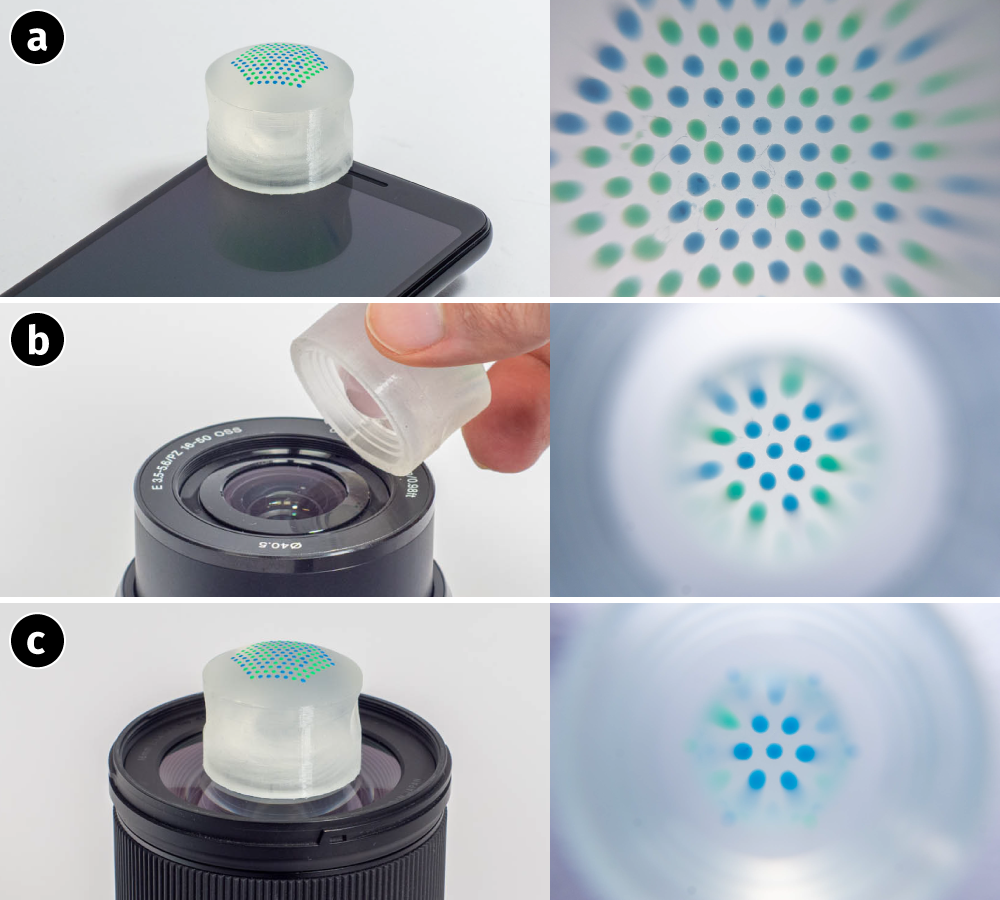}
    \caption{Three lenses with an identical field of view of 84\textdegree\ but increasing entrance pupil diameters: a) Google Pixel 3a front camera b) Sony SEL-P1650 lens (16mm focal length) c) Sigma 16mm 1.4 DC DN (16mm focal length). For all three images, the LensLeech is resting directly on the front element of the camera lens.}
    \label{fig:entrancepupilsizes}
\end{figure}

The main limitation when using any optical attachment on lenses is the entrance pupil diameter and its distance from the first surface of the lens.
The entrance pupil is a virtual opening within the lens barrel through which all entering light rays pass. Size and position within lenses can vary across lens designs, even when an image of a distant object taken with different lenses would look identical (see fig. \ref{fig:entrancepupilsizes}).
The silicone attachment (in the size as presented) works well on small and medium-sized lenses but requires a different geometry on very large lenses such as professional photography or videography lenses with large front elements and entrance pupils for better low-light performance. As a rule of thumb: if the image of the aperture seen through the front element of the lens is considerably larger than the silicone lens (12mm in diameter) the number of visible points is strongly reduced. In general, a lower bound of 19 points is required to reliably recognize input gestures through the soft widget.
Additional limiting factors on the optical system are the field of view of the lens and the curvature of the first glass element of the lens. 
A camera with a narrow field of view will reduce the number of visible points, similar to a large entrance pupil. This makes the presented concept more suitable for medium to wide-angle systems such as webcams, smart home devices, smartphones, and wearable cameras.
If the LensLeech is used with hard attachments (such as the lens cap) it does not sit directly on the glass and a strong lens curvature is not an issue. 

\section{Discussion}\label{section:discussion}

\begin{figure}
    \centering
    \includegraphics[width=\linewidth]{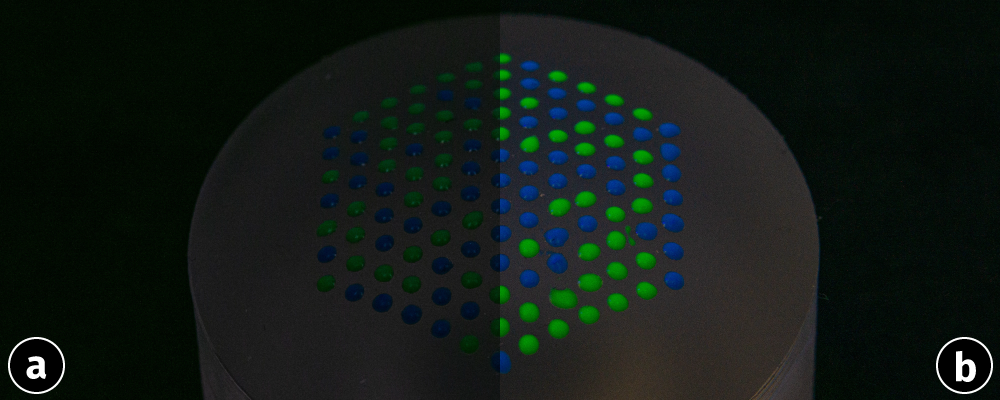}
    \caption{Near-UV flood illumination can considerably increase the brightness and contrast of the UV-reactive pigments in the point pattern while it only marginally brightens the environment. a) No ultraviolet illumination b) Single 365nm-wavelength light source.}
    \label{fig:uvillumination}
\end{figure}

Compared to other on-lens interaction concepts such as \textit{CamTrackPoint} ~\cite{yamadaCamTrackPointCameraBasedPointing2018} and \textit{LensGestures} ~\cite{xiaoLensGestureAugmentingMobile2013}) that process unfocused light, the LensLeech is less limited in the amount of information it provides but it requires ambient light as well. While the LensLeech performs well in most situations, LEDs (such as smartphone flashlights and autofocus-assist lights of still cameras) or screens of devices can be used to provide additional artificial illumination. This can be seen in the hybrid viewfinder: a section of the covered display illuminates the point pattern from below. Depending on the device and application scenario this might not be a viable option. For usage within a predefined space, near-ultraviolet flood lights can be installed to brighten the UV-reactive pigments in the point pattern (see fig. ~\ref{fig:uvillumination}) with little interference to the brightness of the environment.

In general, an attachment solely made from silicone is simple, robust, and---to an extent---expendable. Material cost per piece is about 2 USD/EUR when fabricated in small quantities. This makes the LensLeech comparable to other inexpensive attachments for mobile devices that extend I/O capabilities, such as Google Cardboard ~\cite{googlecardboard} or Nintendo Labo ~\cite{nintendolabo}. Similar to the limited lifetime of corrugated cardboard, a LensLeech and its point pattern may eventually suffer from wear and tear after extended usage. 

A possible negative perception of letting an object touch the front element of the lens is not an issue when used on smartphone lenses. The application examples show that in other scenarios it makes sense to use device-dependent additions such as a rigid shoe on protruding lens barrels for action cameras or lens caps on interchangeable-lens cameras.

\section{Future Work}

A LensLeech is uniformly made from a single silicone formula with consistent Shore hardness throughout the whole body. By making use of a two-stage mold, the lower part of the body containing the lens could be molded separately with a harder type of clear silicone, resulting in a lower deformation of the lens when compressed. Also by integrating air-filled cavities and compliant elements in multi-stage molds, tactile feedback can be provided, resulting in a sensation when a certain amount of force is applied. 

The limitation of large entrance pupil sizes can be circumvented by replacing the single silicone lens with a grid of smaller lenses. This requires a different fabrication technique for the mold and a point pattern that is aligned with camera lens angle and microlens position. This limits the silicone attachment compatibility to only a single lens, yet this may not be an issue for applications such as model-specific lens caps.
While only a single type of LensLeech is presented, the concept is versatile. With additional illumination, microlens arrays would allow emulating a small touchscreen on the top surface while using an angled surface geometry makes sensing fingerprints possible.
In the near future, the emergence of cameras under displays in smartphones would allow the use of silicone attachments as tangible input and output devices in tabletop-like scenarios. The display can be used both for illuminating the LensLeech to sense in dark spaces as well as to display output in or on the body itself by refracting and redirecting the light. 

\section{Conclusion}

We presented the LensLeech, a soft silicone attachment that allows to sense pushing, pressing, rotating, and squeezing when placed directly on or above lenses of arbitrary cameras. This makes it possible to add tangible input methods to a wide range of existing and new devices, especially small action or lifelogging cameras and smartphones. We have shown application examples ranging from small, body-worn devices to lens caps for large cameras and complex smartphone attachments.
While the attachments are limited in their compatibility mainly by lens geometry, the low-light performance allows them to be used with only ambient illumination without any need for hardware modifications on a wide range of existing devices.
This simple and inexpensive approach opens up an interaction space on lenses for rich input that was previously inaccessible with a range of further applications in the (soft) robotics domain.

\section*{Reproduction Note}\label{section:reproductionnote}

The example applications, source code, CAD models of molds and fixtures, a detailed description of the fabrication process, and data/scripts for generating plots are available publicly:\newline
\mbox{\url{https://github.com/volzotan/LensLeech}}

\begin{acks}
This work was funded by the Deutsche Forschungsgemeinschaft (DFG, German Research Foundation) through project EC437/1-1.
\end{acks}

\bibliographystyle{ACM-Reference-Format}
\bibliography{bibliography,bibliography_manual}

\end{document}